\documentclass[conference]{IEEEtran}
\IEEEoverridecommandlockouts
\usepackage{cite}
\usepackage{graphicx} 
\usepackage{xurl}
\usepackage{amsmath,amssymb,amsfonts}
\usepackage{textcomp}
\usepackage{hyperref}
\usepackage{listings}
\usepackage{xcolor}
\def\BibTeX{{\rm B\kern-.05em{\sc i\kern-.025em b}\kern-.08em
    T\kern-.1667em\lower.7ex\hbox{E}\kern-.125emX}}

\begin{document}
\title{The Kubernetes Security Landscape: AI-Driven Insights from Developer Discussions\\
}
\author{\IEEEauthorblockN{J. Alexander Curtis}
\IEEEauthorblockA{\textit{Computer Science Dept.} \\
\textit{Boise State University}\\
Boise, ID. USA \\
alexcurtis@u.boisestate.edu}
\and
\IEEEauthorblockN{Nasir U. Eisty}
\IEEEauthorblockA{\textit{Computer Science Dept.} \\
\textit{Boise State University}\\
Boise, ID. USA \\
nasireisty@boisestate.edu}
}
\date{April 2024}

\lstset{
    frame=single, 
    basicstyle=\ttfamily, 
    keywordstyle=\color{blue}, 
    commentstyle=\color{gray}, 
    stringstyle=\color{red}, 
    breaklines=true, 
    showstringspaces=false, 
    captionpos=b, 
    escapeinside={(*@}{@*)}, 
}



\maketitle

\begin{abstract}
\textit{Context}: 
Kubernetes, the go-to container orchestration solution, has swiftly become the industry standard for managing containers at scale in production environments. Its widespread adoption, particularly in large organizations, has elevated its profile and made it a prime target for security concerns.
\textit{Objective}: 
This study aims to understand how prevalent security concerns are among Kubernetes practitioners by analyzing all Kubernetes posts made on Stack Overflow over the past four years. 
\textit{Method}: 
We gathered security insights from Kubernetes practitioners and transformed the data through machine learning algorithms for cleaning and topic clustering. Subsequently, we used advanced AI tools to automatically generate topic descriptions, thereby reducing the analysis process.
\textit{Results}: 
In our analysis, security-related posts ranked as the fourth most prevalent topic in these forums, comprising 12.3\% of the overall discussions. Furthermore, the findings indicated that although the frequency of security discussions has remained constant, their popularity and influence have experienced significant growth.
\textit{Conclusions}: 
Kubernetes users consistently prioritize security topics, and the rising popularity of security posts reflects a growing interest and concern for maintaining secure Kubernetes clusters. The findings underscore key security issues that warrant further research and the development of additional tools to resolve them.

\end{abstract}


\section{Introduction}\label{sec:introduction}

Kubernetes is an industry-leading orchestration platform with unprecedented adoption across organizations from startups to Fortune 100 companies~\cite{kubernetesCaseStudies}. As a container orchestration platform, Kubernetes enables the management of advanced applications on a massive scale.
As the growth of Kubernetes has reached the mainstream, it is becoming more critical than ever to understand this platform and its security posture. Security has been identified as the \#1 IT-funded priority for 2023, and security is the biggest concern with Kubernetes adoption~\cite{redhatGlobalCustomer}. 

Kubernetes' unrivaled capability comes with the trade-off of complexity. According to a recent survey of 600 engineers, DevOps, and security professionals, more than 50\% stated that they were concerned about misconfigurations and vulnerabilities in their Kubernetes clusters~\cite{Red_Hat2023-bz}. These concerns are reasonable due to the nature of Kubernetes being remarkably dynamic and highly complex, comprising numerous components that must all have precise complementing security configurations.
Some work has already been done to analyze security misconfigurations and weaknesses through repository mining~\cite{Rahman2023-xm}. This work provides insight into existing open-source configurations. 
However, the motivation of our paper is to understand the attitude of developers and discussions around Kubernetes security to see which topics receive the most attention from developers and which areas need additional clarity and improvement.

Stack Overflow acts as the center of community for many technology tools. Engineers use it to post questions, promote discussion, and build community around specific technologies. The site offers a wealth of information about what developers are discussing, including difficulties or limitations of existing technologies. We perceive these discussions as a proxy for quantifying the current needs and concerns of Kubernetes development.
Using text extraction and Machine Learning, incorporated with new AI tools, we have used this platform to analyze the current landscape of developer sentiment and Kubernetes discussions. 

By analyzing trends over time, we can also understand, using practical metrics, whether Kubernetes' security posture is improving or degrading. We performed this analysis by seeing the number of new questions being asked, which could indicate an increase in new concerns. Similarly, by comparing view analysis over time, we can adequately estimate the growth rate of Kubernetes to further quantify its adoption.

Therefore, we pose the following research questions to improve the knowledge landscape of Kubernetes security.

\begin{itemize}
    \item{\textbf{RQ1: What types of Kubernetes posts are developers actively discussing on community channels?}}
    \item{\textbf{RQ2: What Kubernetes security concerns receive the most attention or discussion from users?}}
    \item{\textbf{RQ3: What ratio of Kubernetes security concerns get answered or resolved by the community?}}
    \item{\textbf{RQ4: How has the quantity of security posts evolved over time?}}
\end{itemize}


The results of this work can inform researchers about areas in the Kubernetes landscape that are misunderstood or underdeveloped. Our results will also be effective in directing researchers, developers, and maintainers of the Kubernetes project about areas to focus on in future platform development and feature exploration.

Our work aims to map the landscape of Kubernetes practitioner interest, focusing on security concerns to guide future research, evolution, development, and enhancements by identifying critical areas needing attention and improvement.

In summary, we make the following key contributions:

\begin{itemize}
    \item A cleaned and standardized dataset that contains all 35,417 Kubernetes Stack Overflow posts from the past four years.
    \item A quantitative analysis of the top security topics that are discussed in community forums with comparative ranking by various popularity metrics to suggest areas of Kubernetes that require additional training, documentation, tooling, and research.
    \item An analysis of the effectiveness of using new Large Language Models to help researchers with categorization and summarization tasks.
\end{itemize}

\section{Background}\label{sec:background}
%

Kubernetes is an open-source container orchestration platform that has emerged as the dominant standard for automating the deployment, scaling, and management of containerized applications and services~\cite{Islam_Shamim2020-cl}. Kubernetes has coincided with the rise of cloud computing, as the container orchestration model enables the elastic computing platforms that users take for granted today~\cite{Medel2018-rf}. Kubernetes addresses the complexity of deploying and managing a software system by modularizing business logic. It introduces a variety of primitives and abstract concepts\footnote{Examples of these primitives and abstractions include: ConfigMaps, Deployments, Labels, Namespaces, Pods, ReplicaSets, Secrets, Selectors, Services, StatefulSets, and Volumes.} for distributed applications, offering an object-oriented approach to building distributed applications~\cite{Jiao2021-ue}.

Google has been running containerized workloads in production for more than two decades, starting with the introduction of Borg and Omega to automate the management and scaling of their applications~\cite{Burns2016-rr}. Kubernetes was created at Google in 2014 as a re-build of these earlier systems to improve on their weaknesses and reinforce their strengths~\cite{Burns2016-rr}. Kubernetes was officially released as an open-source project in July 2014, and in 2015, Google partnered with the Linux Foundation to form the Cloud Native Computing Foundation (CNCF), where Kubernetes became its first project~\cite{cncf-whoweare}.

The CNCF plays a pivotal role in guiding and supporting the development of cloud-native applications and technologies. In 2022, the CNCF reported that 96\% of companies are either using or evaluating Kubernetes~\cite{cncf-2022}. More than half of Fortune 100 companies have already adopted Kubernetes as their orchestration platform. In addition, 78\% of small and medium-sized businesses currently use Kubernetes~\cite{Red_Hat2023-bz}. 

The market share of Kubernetes is expected to grow remarkably in the upcoming years. Projections show an increase from its current state at USD \$1.8 Billion in 2022 to a dominating USD \$7.8 Billion by 2030~\cite{globenewswireLatestGlobal}. This industry projection represents an expected Compound Annual Growth Rate (CAGR) of 23.40\%. These numbers only represent the market for Kubernetes services but don't scratch the surface of the overall financial impact that Kubernetes powers for companies through products and services that are maintained by Kubernetes as a platform. Kubernetes is relied on by everyone from tech giants such as IBM and Huawei, financial institutions such as BlackRock and CapitalOne, consumer products such as Spotify and Booking.com, and government organizations such as NASA and the city of Montreal, to name a few~\cite{kubernetesCaseStudies}. It is safe to say that most of the technology services that you use today are powered by Kubernetes, making it one of the most prevalent tools in the technology industry. 

As Kubernetes grows in its adoption to a ubiquitous level, the importance of maintaining the security of the platform becomes a larger target and of paramount concern. Companies like Uber~\cite{uber_microservice} and Netflix~\cite{netflix_microservices} each use Kubernetes to manage hundreds of thousands of containers simultaneously and keep these always-available systems online~\cite{Blaise2022-ia}. 

The implications of a security risk in Kubernetes can be disastrous since clusters have ``ultimate power'' over an organization's environment. For example, in 2018, attackers gained access to Tesla's Kubernetes cluster, which enabled them to deploy new compute instances in Tesla's cluster to mine cryptocurrency for attackers~\cite{securitybriefHackersExploit}. The same attackers were also able to gain access to administrative credentials for Tesla's cloud provider (AWS), where they could download files stored in S3. A 2023 survey of 800 security and IT leaders from large organizations in the US, UK, France, and Germany found that 59\% have experienced security incidents in their Kubernetes environments~\cite{infosecuritymagazineOverHalf}. 

In 2024, a vulnerability known as \textit{Sys:All Loophole} was discovered that allowed an attacker to take over a Google Kubernetes Engine (GKE) cluster \textbf{with any Google account} due to a simple misconfiguration~\cite{orcaSysAllOverview}. This vulnerability was found to be active in more than 250,000 production Kubernetes clusters, including at least one major NASDAQ-listed corporation~\cite{orcaSysAllNasdaq}. In the case of the NASDAQ-listed company, a security research group was able to gain administrative access to the company's cloud accounts in both AWS and GCP. They also demonstrated the ability to download Google Oauth credentials, sensitive company files from S3 buckets, and private encryption keys. Furthermore, they gained administrative access to the company's RabbitMQ cluster, Elastic account, and private container registries~\cite{orcaSysAllNasdaq}. This system-wide security breach caused by a single Kubernetes vulnerability highlights the importance of security in this platform.

A limitation of current research is that all security analyses have been performed on open-source projects. Although Stack Overflow is \textit{also} open information, it contains discussions related to questions and situations encountered by developers while using both open-source and closed-source projects. We expect Stack Overflow to provide a more holistic perspective on Kubernetes' security concerns than analyzing open-source projects alone.

\section{Methodology}\label{sec:methodology}

\begin{figure*}
    \centering
    \includegraphics[width=0.75\linewidth]{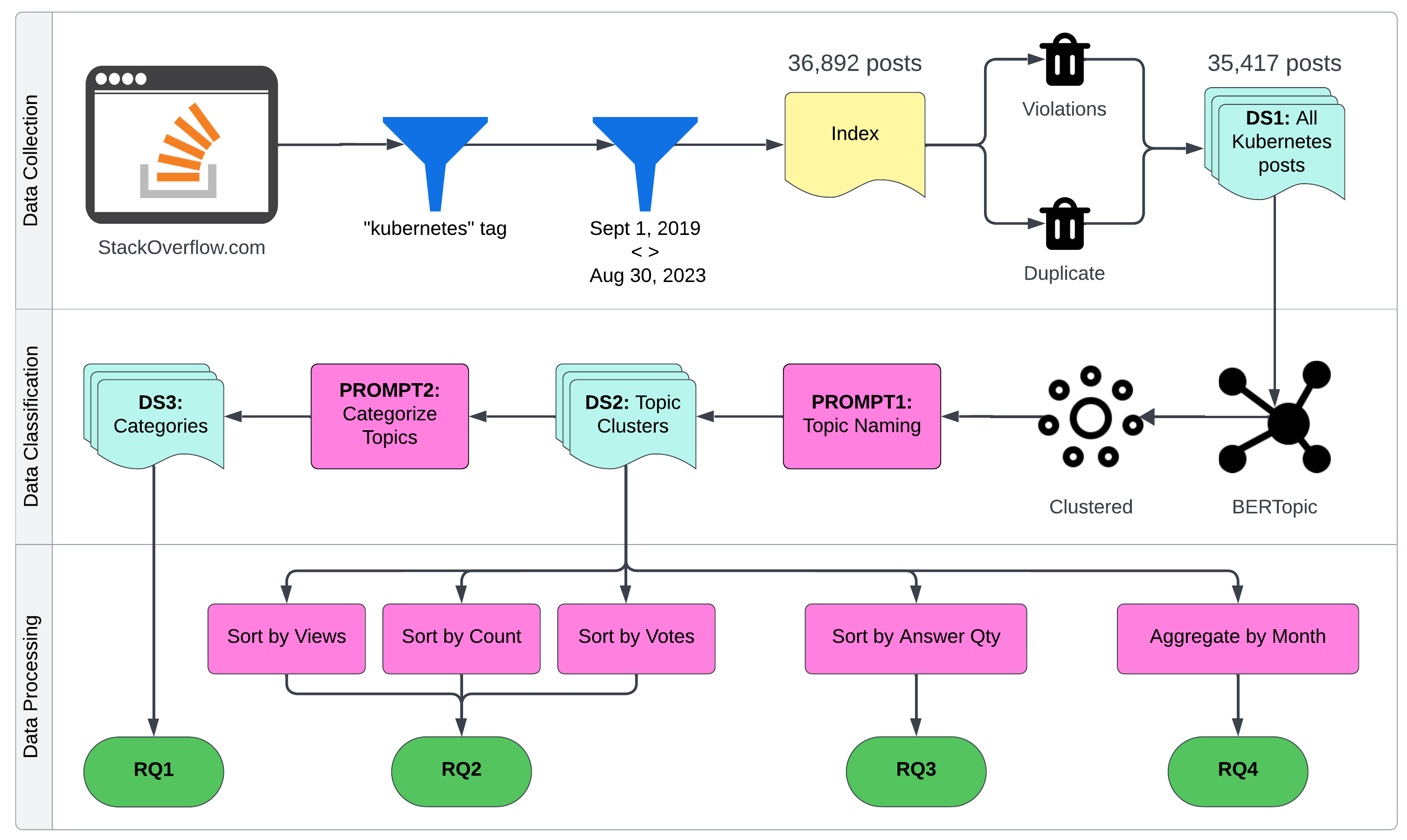}
    \caption{Visualization of the Research Methodology}
    \label{fig:methodology-flowchart}
\end{figure*}

This study methodology can be broken down into three phases: {C}ollection, {C}lassification, and {P}rocessing. These phases are visualized with Figure~\ref{fig:methodology-flowchart} and are addressed individually in the following text.

\subsection{Data Collection}

In an effort to gain an understanding of developer concerns and grasp the current landscape of Kubernetes, we leveraged the data available on Stack Overflow as a way to answer the research questions. We decided to use Stack Overflow as the source for capturing the state of community questions in this study because of its prevalence as the top developer resource when discussing technology problems.

Stack Overflow contains questions on nearly every topic of software engineering. The site implements a tagging system that is used to organize and discover posts. Each post must have at least one tag, with a maximum of five~\cite{stackoverflowBestPractices}. We use the tag \textit{kubernetes} to filter all posts related to the Kubernetes platform.
Stack Overflow moderators that specialize in Kubernetes manually maintain the accuracy of the tagging on Kubernetes posts, both adding and removing the tag on relevant and irrelevant posts as they are created and answered. 

\subsubsection{\textbf{Web Scraping}}
For the data collection task, we created a Python web scraper to collect data for all Stack Overflow posts tagged with the \textit{kubernetes} tag~\cite{stackoverflowkubernetesSynonyms}. To ensure that these posts are recent and relevant, we only collected posts that had been updated in the past four years. The effective dates allowed in the dataset are all posts with an \texttt{updated\_at} field between September 1, 2019 and August 29, 2023. We used the \texttt{updated\_at} field instead of \texttt{created\_at} to allow older posts with recent updates to be included in the dataset, as they are still gaining attention from the community. 

The scraping process was carried out in two stages. First, we collected an index for every post tagged as \textit{kubernetes}. Subsequently, we performed a detailed scrape on each individual page from the index to collect detailed data for each post. During the detailed scrape, any post that was marked as duplicate or had been reported for community violations on Stack Overflow was removed from our dataset. The initial collection of \textit{kubernetes} tagged posts between the effective dates was 36,892 posts. After the removal of duplicates and community violations, there were 35,417 unique posts. These posts become the dataset used for data classification known as \textbf{DS1} (as seen in Figure~\ref{fig:methodology-flowchart}) containing all Kubernetes posts matching our collection criteria.

The data collected for each post includes:

\begin{itemize}
    \item \textbf{id}: Unique identifier for post. Grabbed from url and determined by Stack Overflow
    \item \textbf{title}: The post title
    \item \textbf{url}: The url for the original post on Stack Overflow
    \item \textbf{tag\_array}: All tags for the post. One of which will always be \textit{kubernetes}, but other tags may exist~\cite{stackoverflowBestPractices}.
    \item \textbf{content}: Text content with HTML markup removed
    \item \textbf{votes}: Net number of votes for the post when scraped.\footnote{Posts can receive positive and negative votes, raising and lowering this number. Total votes can be negative.}
    \item \textbf{answers}: Total number of solutions provided by the community members to the post's question.
    \item \textbf{views}: Total number of public views the post has received at the time of scraping, as reported by Stack Overflow.
    \item \textbf{accepted}: Boolean field indicating whether there was an answer that was marked as ``accepted.''
    \item \textbf{definitive}: Boolean field indicating whether Kubernetes moderators have marked the post as ``definitive,''.\footnote{Posts marked as ``definitive'' are considered to be common questions with high-quality responses. These are updated frequently and treated like a wiki page. New questions created that are considered duplicates of a pre-existing definitive post will be marked as duplicate by moderators and get redirected to the relevant definitive post.}
    \item \textbf{created\_at}: Timestamp of post creation.
    \item \textbf{updated\_at}: Timestamp of last post update. \footnote{\texttt{updated\_at} changes could be from the moderators or the creator of the post indicating changes to the original post title and content. Answers do not alter the timestamp.}
\end{itemize}

\subsubsection{\textbf{Data Cleaning}}
In later steps, we performed topic clustering. Due to the boilerplate sample code submitted in many Stack Overflow posts, code snippets have been known to reduce the accuracy of clustering. Therefore, we decided to remove the code from the posts' content before it is fed into clustering algorithms. Data cleaning allows the clustering algorithm to focus on the content of the post rather than being thrown off by words in the code, which might be similar even for very different topics~\cite{Ramos2003-rd}. We cleaned our data by removing all code snippets from the content of each post. Then, we stored the cleaned data as an additional field called \texttt{content\_clean} in the database. We also stored the original content for later comparison or recollection.


We stored the collected data in a MySQL database, facilitating easy querying, updating, and exporting of the dataset at various stages of the research. We provided and made available the SQL dump of our data as part of this paper.

\begin{lstlisting}[caption=GPT-4 example of PROMPT1 used for topic determination, label=lst:gpt-topic]
(*@{\textbf{System:}}@*) You are a helpful assistant who understands kubernetes. You are helping a user who is trying to understand a topic based on a list of keywords. You must provide a short summary of the topic based on keywords. Be as concise as possible, using between 1-5 words. Topics do not need to be full sentences. Respond with only the topic name. For example if the keywords are "certificate, https, ssl, certmanager, certificates, tls, cert, ingress, ca, http", you could respond with: TLS/SSL Certificate Management.
(*@{\textbf{Prompt:}}@*) Topic words: dns, coredns, resolve, nslookup, etcresolveconf
(*@{\textbf{Response:}}@*) DNS Resolution in Kubernetes
\end{lstlisting}

\subsection{Data Classification}

After collecting and storing the data in a database, we first clustered them, then classified each cluster with a topic name, and finally grouped topics into broad categories for analysis.

\subsubsection{\textbf{Clustering}}

There are a total of 35,417 posts in the dataset, known as \textbf{DS1} in Figure~\ref{fig:methodology-flowchart}. Although each post is unique, many of these posts are related to similar topics. To answer the research questions, we must determine similar posts and aggregate them for quantitative analysis. We used c-TF-IDF clustering method~\cite{Grootendorst2022-yk} to find similar posts based on relationships and commonalities of terms used within the post. This method has been found by prior work to be the most effective way to categorize and group Stack Overflow posts~\cite{Ponzanelli2014Improving, Yazdaninia2021Characterization, Egger2022-qo}.

The tool we used to implement this algorithm is BERTopic~\cite{github_bertopic}, which provides a framework for combining different pre-processing steps and customizing the algorithm steps for better accuracy~\cite{Grootendorst2022-yk}.


For the purposes of this study, we define topics and categories as follows:

\begin{itemize}
    \item \textbf{Category:} A broad subject of the given cluster.
    \item \textbf{Topics:} A specific subject or discussion within a category.
\end{itemize}

In summary, \textbf{categories} are broad sections used to organize content into general themes, while \textbf{topics} are specific subjects within those categories that address particular questions, discussions, or areas of interest. A category has many topics, but a topic has only one category. If a topic relates to more than one category, we determined and used the strongest relationship.

Other than the initial \textit{kubernetes} tag, which we used for scraping, all of the topics and categories in this study were determined by us and not Stack Overflow. Due to the high volume of posts (35,417), we decided to leverage a large language model (LLM) to identify categories and topics.

\begin{lstlisting}[caption=GPT-4 example of PROMPT2 used for category determination, label=lst:gpt-category]
(*@{\textbf{System:}}@*) I have chosen 10 categories:  "Deployment CI/CD", "Security", "Networking", "Storage", "Monitoring & Logging", "Application/Microservices/Jobs/Messaging", "Containerization", "Cluster Configuration", "Cloud Services", "Scaling & Load Balancing", "Miscellaneous/ Other". For each prompt I will provide you with a topic and you must choose the most appropriate category that the topic belongs to. If it fits into more than one, then choose the strongest fit. Only select the miscellaneous & other category if it does not strongly fit into any other topic. Respond with the selected category followed by a colon. Then provide an explanation for why that category was selected on the next line.
(*@{\textbf{Prompt:}}@*) TLS/SSL Certificate Management
(*@{\textbf{Response:}}@*) Security: TLS/SSL certificate management is primarily concerned with the encryption and secure communication between clients and servers, which falls under the broader category of cybersecurity and protection mechanisms, thus aligning with the Security category.
\end{lstlisting}

\subsubsection{\textbf{Topic Determination}}

The BERTopic processing groups various related posts with each other as a cluster. However, this process does not give these clusters a topic name. Instead, the output provides the common terms found within that cluster. By observing these keywords, it is possible to define a human-readable topic name to represent each cluster. The topic name would describe all the posts within that cluster. We used an LLM to define a topic name based on these keywords. We sent the dataset, which includes each cluster along with its keywords, to the OpenAI API (utilizing the GPT-4 model) in order to generate a topic label for each cluster~\cite{Liu2023-ql}. The prompt used to generate these topics is shown with an example in Listing~\ref{lst:gpt-topic}. This step is also represented as \textbf{PROMPT1} in Figure~\ref{fig:methodology-flowchart}.

\subsubsection{\textbf{Category Determination}\label{sec:data-classification-categories}}

Each topic must also be grouped into a broad category. We defined the most common categories to which all Kubernetes posts could belong. We determined the following set:

\begin{itemize}
    \item Deployment CI/CD
    \item Security
    \item Networking
    \item Storage
    \item Monitoring \& Logging
    \item Application Layer
    \item Containerization
    \item Cluster Configuration
    \item Cloud Services
    \item Scaling \& Load Balancing
\end{itemize}

We used the LLM model again, asking the model to determine which category was the best fit for each given topic~\cite{Liu2023-ql}. The prompts and an example are provided in Listing~\ref{lst:gpt-category} to demonstrate how a category was determined. This step is represented as \textbf{PROMPT2} in Figure~\ref{fig:methodology-flowchart}. We manually validated the top 50 posts to confirm accuracy and we were satisfied with the LLM's results, finding that it categorized the topics correctly across all of the top posts.

\subsection{Data Processing}

Once the collected and classified data is available, we are ready to analyze it to address our research inquiries. The methodology for data processing varies according to each research question. The following sections will detail the procedures employed to address each research question.



\subsubsection{\textbf{RQ1: What types of Kubernetes posts are developers actively discussing on community channels?}}

The goal of this question is to understand what common types of posts are being asked by Kubernetes Practitioners. This insight will help us better understand common patterns and areas of Kubernetes administration that are confusing to practitioners. RQ1 will inform us about the general interests or needs of the Kubernetes community in relation to other posts. We are particularly interested in the security category. In the final three research questions, we will analyze the posts within the security category.

The categories were classified in Section~\ref{sec:data-classification-categories}, and this is the same categorization data we will use.

\subsubsection{\textbf{RQ2: What types of Kubernetes security concerns receive the most attention?}}

An area of particular interest for us is understanding the security landscape of Kubernetes. By diving deeper into the Stack Overflow posts categorized under \textit{Security} (from RQ1), we can use the topic clusters to identify common security issues that are being discussed within the Kubernetes community. RQ2 creates a taxonomy of security-related topics from the \textit{Security} category of RQ1. With this data, we seek to discover which security topics receive the most attention.

To answer this question, we extracted topics that were in the \textit{Security} category determined in RQ1. This extraction isolates the security topics to become the dataset \textbf{DS2}. The security topics will be ranked according to three different metrics: views, answers, and comments. These standard indicators of engagement, each offer different insights into Kubernetes security topics~\cite{Barua2014-di, Choetkiertikul2015-wz, Bhat2014-zb}. The results also offer a comparative analysis of which security topics are popular or the most misunderstood.

\subsubsection{\textbf{RQ3: What ratio of Kubernetes security concerns get answered?}}

The next area of interest is exploring the percentage of Kubernetes security questions that are getting answered by the community.

We answer this question using \textbf{DS2}, similar to RQ2. However, this time, we compared the cumulative number of posts with accepted answers to the unanswered posts as a ratio. These results allow us to understand which security topics go unanswered, possibly indicating areas that require future evolution for the platform. The results can also indicate topics where community education should be improved, such as the addition of better documentation or training.

\subsubsection{\textbf{RQ4: How has the quantity of security posts evolved over time?}}

Our final question seeks a time-based analysis of how different security posts have evolved over time. 

To gain this analysis, we used the category dataset \textbf{DS3} filtering only the posts marked as \textit{Security} and plotted the posts' creation over time. To do this, we aggregated the number of Kubernetes security posts into periods of months when the post was published. Then, we plotted the sum of posts published each month in a timeline. Finally, we performed a trend analysis to see if security posts have historically increased, decreased, or remained stable.

This question provides insight into whether security posts are becoming more common or less common to determine whether security issues are becoming more prevalent or awareness has improved.

\section{Results \& Discussion} \label{sec:results}

We provide answers and implications to our research questions in the following sections.

\subsection{\textbf{RQ1: What types of Kubernetes posts are developers actively discussing on community channels?}}\label{sec:results-rq1}

The motivation of this question is to provide insight into areas of shared confusion or difficulty that Kubernetes practitioners face while working with the platform. From that insight, we want to know if questions categorized as \textit{Security} are gaining a significant share of the total questions on Stack Overflow. These findings indicate to platform developers the rate at which security concerns affect practitioners compared to other types of platform concerns. 

\begin{figure}
    \centering
    \includegraphics[width=1\linewidth]{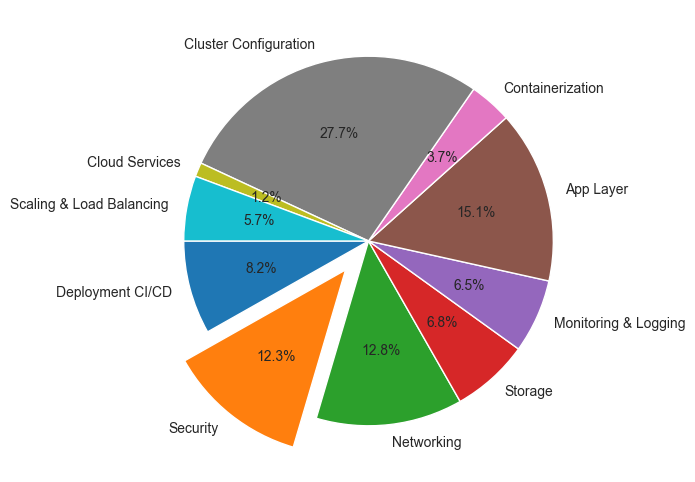}
    \caption{Distribution of Question Types by Category}
    \label{fig:question-distribution}
\end{figure}

Of the 35,416 posts that were extracted from Stack Overflow, the total corpus (group of posts) was clustered by BERTopic. The BERTopic ML algorithm defined related posts as clusters. Each cluster represents a set of posts that are considered similar to each other. The 35,416 individual posts were distilled into 273 unique topics. These topics provide a detailed summary of many similar questions, making it easier to analyze the types of questions being asked on Stack Overflow.

For example, 1,335 questions related to \textit{Ingress Path Configuration}. So, while many different posts have been created that ask slightly different specific questions, all of these posts are related to the topic of configuring ingress paths within the cluster. These topics will be useful for our more in-depth analysis performed in RQ2. However, we want to get a macro view of questions from broad categories. For the example of \textit{Ingress Path Configuration} topic, we could categorize this topic under the \textit{Networking} category. Likewise, 427 documents ask questions about DNS Resolution, which we could also categorize as \textit{Networking}.

We selected 10 broad categories to use to analyze our data. This selection will give us a macro overview of the general types of questions that get asked most often on Stack Overflow. The 10 categories we chose are:

\begin{enumerate}
    \item \textbf{Deployment CI/CD}: DevOps, deployments, Helm\footnote{\textit{Helm}, which is a very popular Kubernetes package manager to simplify deployment across environments}, \mbox{ArgoCD}, and pipelines.
    \item \textbf{Security}: Secrets, Certificates, Encryption, exploits, security breaches, etc
    \item \textbf{Networking}: Service meshes, IP addressing, Ingress, and other networking concerns.
    \item \textbf{Storage}: Persistent storage, volumes, \texttt{etcd}\footnote{\texttt{etcd} is a distributed persistent data store which is core to Kubernetes clusters as it stores the deployed state of applications, values for secrets, and resource configuration. It is pronounced (et-see-dee)}, etc.
    \item \textbf{Monitoring \& Logging}: Observability, logging frameworks (DataDog, fluentd, Logz.io, etc), APM, etc.
    \item \textbf{Application Layer}: Anything related to getting the application running inside of Kubernetes. These posts could include configuring a Spring Boot application, configuring an Airflow cluster, Apache Spark jobs, and everything in between. If it was application-specific (and not related to Kubernetes itself), then it was categorized here.
    \item \textbf{Containerization}: Docker, \texttt{containerd}, and other CRIs. Any troubleshooting that is specific to building or running the container.
    \item \textbf{Cluster Configuration}: Core Kubernetes questions about setup, configuration, settings, and so forth. This category may also contain questions about configuring OpenShift and other implementations of Kubernetes.
    \item \textbf{Cloud Services}: Posts specific to a cloud provider like AWS. This category includes setup configuration for cloud-specific implementations like EKS\footnote{EKS is Amazon AWS' \textit{Elastic Kubernetes Service}. A hosted Kubernetes product}, GKS\footnote{GKS is Google Compute Cloud's \textit{Google Kubernetes Service}. A hosted Kubernetes product}, and AKS\footnote{AKS is Microsoft Azure's \textit{Azure Kubernetes Service}. A hosted Kubernetes product}.
    \item \textbf{Scaling \& Load Balancing}: Autoscaling, Horizontal Pod Autoscaler, Load Balancers, etc.
\end{enumerate}

Each of the 273 topics was placed into one of the 10 categories listed above. We present the results in Fig.~\ref{fig:question-distribution}.

The most common question category is \textit{Cluster Configuration}. This category includes general questions about setting up and configuring a Kubernetes cluster. It accounts for 27.7\% of all questions posted. This result seems reasonable since these types of questions involve tasks that everyone must do in order to start working with Kubernetes or setting up a new cluster. Kubernetes is also very complicated, resulting in many settings and configurations that need to be set precisely. Therefore, we expected to see a large number of questions about cluster configuration.

The second largest category is questions about the \textit{Application Layer}. This category includes questions related to problems getting a specific application working within Kubernetes.

The third and fourth most common questions are \textit{Networking} and \textit{Security} at 12.8\% and 12.3\%, respectively. These two categories are statistically equal, which is interesting. In particular, there is a lot of overlap between these two categories. Many networking questions are also security questions.

\subsection{\textbf{RQ2: Which Kubernetes security concerns receive the most attention or discussion from users?}}

To answer this question, we took all topics that were categorized under \textit{security} by our AI-driven categorization in RQ1. We then filter and sort these topics by three attributes: most views (Table~\ref{tab:top-viewed}), most upvotes (Table~\ref{tab:top-voted}), and most posts (Table~\ref{tab:top-counts}).

\begin{table}[h]
    \centering
    \begin{tabular}{| p{6cm} | c |}
         \hline
         TLS/SSL Certificate Management& 2,090,854\\
         \hline
         Secrets Management in Kubernetes& 924,921\\
         \hline
         AWS IAM Roles for EKS& 581,419\\
         \hline
         Kubernetes RBAC Configuration& 516,026\\
         \hline
         File System Permissions& 265,022\\
         \hline
         GCP Identity and Access Management (IAM)& 261,402\\
         \hline
         Keycloak Authentication Flow& 231,640\\
         \hline
         OAuth2 Authentication/Authorization& 205,513\\
         \hline
         Container Security Context& 202,904\\
         \hline
         AKS Identity Management& 141,464\\
         \hline
    \end{tabular}
    \caption{Top 10 Most Viewed Security Topics}
    \label{tab:top-viewed}
\end{table}

\begin{table}[h]
    \centering
    \begin{tabular}{| p{6cm} | p{1.2cm} |}
    \hline
         TLS/SSL Certificate Management.& 1451\\
         \hline
         Secrets Management in Kubernetes& 665\\
         \hline
         AWS IAM Roles for EKS& 417\\
         \hline
         Kubernetes RBAC Configuration& 371\\
         \hline
         GCP Identity and Access Management (IAM)& 280\\
         \hline
         OAuth2 Authentication/Authorization& 169\\
         \hline
         Keycloak Authentication Flow& 163\\
         \hline
         File System Permissions& 159\\
         \hline
         Container Security Context& 123\\
         \hline
         Kubernetes Multi-Tenancy with MinIO.& 92\\
         \hline
    \end{tabular}
    \caption{Top 10 Most Upvoted Security Topics}
    \label{tab:top-voted}
\end{table}

\begin{table}
    \centering
    \begin{tabular}{| p{6cm} | p{1.2cm} |}
    \hline
         TLS/SSL Certificate Management.& 1318\\
         \hline
         Secrets Management in Kubernetes& 661\\
         \hline
         Kubernetes RBAC Configuration& 346\\
         \hline
         AWS IAM Roles for EKS& 239\\
         \hline
         GCP Identity and Access Management (IAM)& 190\\
         \hline
         Keycloak Authentication Flow& 161\\
         \hline
         File System Permissions& 150\\
         \hline
         OAuth2 Authentication/Authorization& 145\\
         \hline
         Container Security Context& 143\\
         \hline
         Azure Kubernetes Service (AKS) Identity Management& 126\\
         \hline
    \end{tabular}
    \caption{Top 10 Most Asked Questions by Security Topic}
    \label{tab:top-counts}
\end{table}

When comparing these results, we find common patterns among all of these ranking factors, with similar topics rising to the top, regardless of which metric is used for ranking. 

The top two topics have such overwhelming popularity that they remain the top two regardless of how we rank them. \textit{TLS/SSL Certificate Management} is overwhelmingly the most popular security topic with 2,090,854 views, 1,451 upvotes, and 1,318 posts created. This is more than double the next ranked topic by every metric. Clearly, this is the most popular topic that users need help with and are interested in discussing.

The second most popular topic is \textit{Secrets Management in Kubernetes}. This topic has 924,921 views, 665 upvotes, and 661 questions asked. Still very dominantly placed, with slightly less than double the third-place topic by most metrics.

The third most popular topic varies slightly based on ranking metrics. However, all results are related to \textit{Identity and Access Management (IAM)}. This group includes AWS IAM, GCP IAM, and Kubernetes RBAC. All of these topics share a similar place in popularity but adjust slightly among each other based on which metric is used. AWS is slightly more popular overall, likely indicative of its dominance as a cloud service provider~\cite{Red_Hat2023-bz}. Kubernetes RBAC is the native IAM implementation within local Kubernetes clusters and also fares strongly against AWS questions, nearly tying it on most measures. It is worth noting that Kubernetes RBAC seems to draw more discussion\footnote{``Discussion'' is determined by increased comments} than AWS IAM. This finding could indicate a lack of clarity when using Kubernetes RBAC since it is drawing more comments with fewer views.

This analysis allows us to determine that the questions that receive the most attention and discussion by developers related to Kubernetes security are (in this order):

\begin{enumerate}
    \item TLS/SSL Certificates
    \item Secrets Management
    \item Identity and Access Management
\end{enumerate}

The ranking of these topics has high confidence, with significant gaps between each topic in terms of popularity, regardless of the metric used to determine popularity.

\subsection{\textbf{RQ3: What ratio of Kubernetes security concerns get resolved by the community?}}

The motivation was to understand whether developers with security questions are getting assistance in securing their clusters. A lack of accepted answers on Stack Overflow for certain security questions identifies an area where further security education or documentation needs to be made available or, furthermore, may be an open concern that requires platform development.

We found that 77.6\% of security questions received at least one published answer, which is consistent with the overall average of 77.8\% of all questions that received at least one answer. This finding indicates that security questions get about the same amount of attention from the community as other questions.

However, only 36.1\% of security questions, on average, had an accepted answer\footnote{An \textit{accepted answer} is one marked by the post author as an answer that they feel answered their question. Some posts may get answers that are not accepted, which generally indicates that the answers were not useful or accurate}. This is slightly below the average of 36.5\% for all questions that receive an accepted answer. This difference suggests a slight hesitance by post authors to accept an answer to security-related questions or them being slightly less satisfied with the answers they received. Although this margin is notably small.

\begin{figure}
    \centering
    \includegraphics[width=1\linewidth]{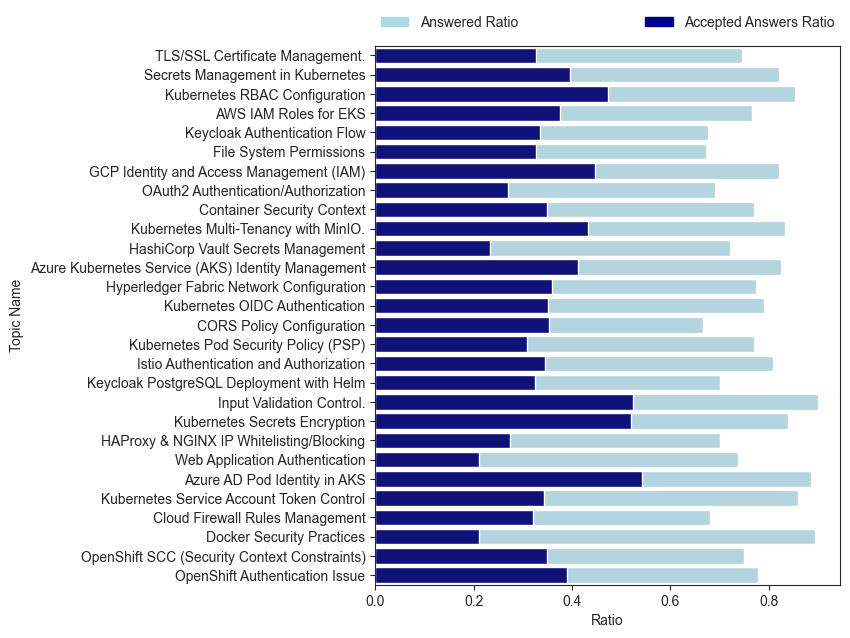}
    \caption{Security Topics based on Ratio of Answers Received}
    \label{fig:answers-ratio}
\end{figure}

Fig.~\ref{fig:answers-ratio} illustrates the ratio of questions that receive any answers (at least one submission) and accepted answers (as indicated by the author of the post) by security topic. One interesting data point is \textit{Docker Security Practices}, which has the highest response ratio, but the lowest accepted response ratio. This discrepancy could be indicative of confusion or disagreement around these security practices, at least in the context of Kubernetes.

We also noted the highest answer ratios from questions regarding access control, both Kubernetes native access control and cloud service Identity and Access Management (IAM). This includes Kubernernetes RBAC (85.2\% answered / 47.4\% accepted), AWS IAM (76.6\% answered / 37.7\% accepted), GCP IAM (82.1\% answered / 44.7\% accepted), and Azure IAM (82.5\% answered / 41.3\% accepted). 

Finally, we wanted to know if there are any questions that receive a high number of views but end up not getting answered. Unanswered posts could be useful in identifying platform shortcomings or unaddressed security weaknesses. 

Looking at questions without answers but with the highest views, we found the following top unanswered security questions:

\begin{flushleft}
\begin{itemize}
    \item \textbf{(5,717 views):} MountVolume.Setup failed for volume "xxx": couldn't get secret
    \item \textbf{(4,035 views):} access pod as root user
    \item \textbf{(3,759 views):} Airflow KubernetesPodOperator : how to access secret passed to Pod?
    \item \textbf{(3,511 views):} Getting \texttt{NET::ERR\_CERT\_AUTHORITY\_INVALID} after switching to another kubernetes cluster
    \item \textbf{(3,436 views):} Getting Unable to connect to the server: x509: certificate is valid for ingress.local, not rancher
    \item \textbf{(3,400 views):} ``secret not found'' reported by ExternalSecret
    \item \textbf{(3,308 views):} Java 8 - \texttt{javax.net.ssl.SSLHandshakeException: Received fatal alert: handshake\_failure}
    \item \textbf{(3,245 views):} Postman \texttt{Error: Exceeded maxRedirects} Probably stuck in a redirect loop after configuring Kubernetes and Ingress Nginx with SSL Certificate
    \item \textbf{(2,603 views):} How to enable login/logout event logging at keycloak with event listener config at Kubernetes?
    \item \textbf{(2,460 views):} Failing to issue TLS certificate with certificate manager in kubernetes \& CloudFlare
    \item \textbf{(2,373 views):} Kubernetes kerberos and kafka secret setup using helm
\end{itemize}
\end{flushleft}

A high number of views is often indicative of the post receiving a lot of search traffic. So, we assume that other users are stuck on these issues, search it on a search engine, and find these posts but are not sure how to get past them. Most of these posts are related to secret management and SSL certificate management. We suggest that these results present areas requiring further platform development and documentation.

\subsection{\textbf{RQ4: How has the quantity of security posts evolved over time?}}

The motivation of this research question is to know whether we are seeing increased activity and interest in Kubernetes security posts over time or whether it has decreased or stayed stagnant. This analysis provides insight into the growth and interest of the platform.

\begin{figure}
    \centering
    \includegraphics[width=1\linewidth]{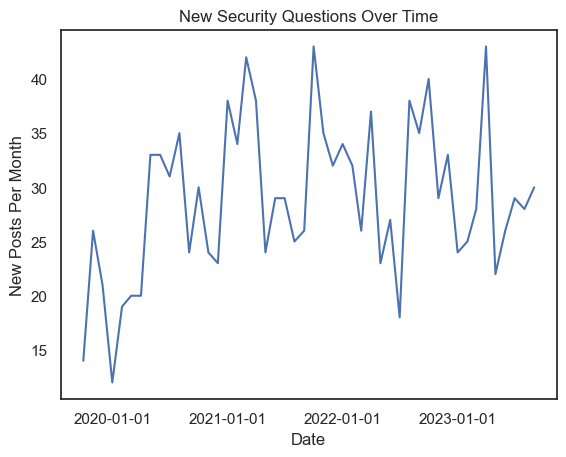}
    \caption{New Kubernetes Security Posts Created Over Time}
    \label{fig:posts-over-time}
\end{figure}

The first results can be seen in Fig.~\ref{fig:posts-over-time}, which shows a slight increase in the number of posts over time but remains relatively stable. The average number of new posts created each month in 2023 was 28.9. This rate equates to 1.45 per business day, when traffic Stack Overflow receives most of its traffic~\cite{StackOveflow-trends2022}. 
This relatively low figure, considering the adoption and market effect that Kubernetes has, could be indicative of a very secure platform.

\begin{figure}
    \centering
    \includegraphics[width=1\linewidth]{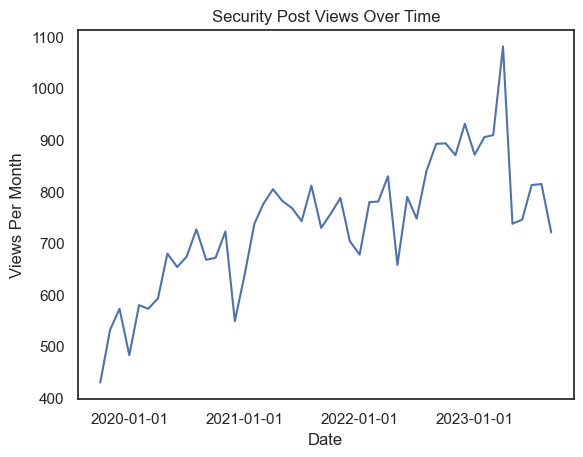}
    \caption{Kubernetes Security Views Over Time}
    \label{fig:views-over-time}
\end{figure}

By contrast, we also wanted to gauge interest in Kubernetes security topics over time. In Fig.~\ref{fig:views-over-time}, we present the views per month on security-related Kubernetes posts over the past four years. The number of views on these questions has increased significantly since 2020 and is showing considerable growth heading into the future.

The contrast between a relatively stagnant new post creation and a significant increase in the number of views suggests the popularity of the platform is rising, while maintaining stable security. The fact that we do not see more security questions being asked as the platform grows could show that all existing questions have previous answers (therefore, there is no reason to make a new question). At the same time, the views increase as more and more organizations adopt the platform and are searching for guidance.

\section{Practical Implications \& Future Works} \label{sec:futurework}

The primary motivation of this paper was to offer an analysis and context for the current landscape of Kubernetes security. We performed this study to encourage and influence the evolution of the platform, which has become so critical to everyday software deployment. In this section, we discuss the practical implications of this study and future works. 

\subsection{Practical Implications}

Here are some practical implications that this paper offers to developers and maintainers of the Kubernetes platform or tools in adjacent areas.

\subsubsection{\textbf{Security should be a top-level consideration in platform maintenance}}

Security interest in Kubernetes is growing over time. As more companies move to Kubernetes, they are considering security as one of their primary requirements. 

\subsubsection{\textbf{Kubernetes platform and tools should be secure by default}}

Misconfigurations exist in over 50\% of kubernetes clusters~\cite{Rahman2023-xm, Mahajan2022-yb, Shamim2021-ti}. Our study provides further confirmation for the frequency of this problem, revealing that around 28\% of all stack overflow discussions are related to configuration. This is an area where developers are the most confused and can lead to major security breaches. Therefore, the configuration options in Kubernetes and related tools must be designed so that they are configured to be secure by default.

\subsubsection{\textbf{The Kubernetes platform is already showing promising growth in improving security}}

One key takeaway from RQ4 was that security posts, which represent new security concerns, are reasonably flat over time. This contrasts with the relative interest that increased significantly over time. The lack of new questions compared to the growth of the platform in recent years indicates a lower question rate per capita. This finding indicates that Kubernetes is already trending in a positive direction in terms of security. 

\subsection{Future Work}

This work should inspire further research in this area. The growing dominance of Kubernetes as a platform encourages the review and improvement of security. This project offers DS1, a cleaned dataset of Kubernetes posts on Stack Overflow with details on content material, views, and acceptance variables. While our paper focused primarily on analyzing the security related posts, future work could dive deeper into other areas like configuration, which accounts for the largest category by questions asked.

Within the topic of security, three areas of greatest interest to Kubernetes practitioners, which also frequently go unanswered, were Certificate Management, Identity Access Management (particularly as it relates to cloud providers), and Secrets Management. These areas lack community consensus and could be improved through further research and exploration.  

\section{Threats to Validity} \label{sec:threats}


Although we made every effort to keep this study as impartial and reproducible as possible, it is necessary to acknowledge a few components of the study.

\subsubsection{\textbf{Internal Threats}} The machine learning algorithm used was BERTopic and is non-deterministic~\cite{Grootendorst2022-yk}, meaning that reproducing the exact results is not always possible. We made multiple passes when running BERTopic to confirm how similar the results were to each other. After analyzing the top 50 results, we found distributions comparable to those shown in Fig.~\ref{fig:question-distribution} across multiple runs. The parameters and libraries that we selected for each step in the pre-processing also have a significant impact on the final results; these were tuned until the results showed high confidence and maintained consistency across multiple runs. It is always possible we did not select the optimal parameters, which could alter the final results.

Our study relied heavily on the use of LLM tools to replace human researchers for topic naming and categorization in our dataset. Using LLMs improved the speed of the analysis but introduced risks. Like ML clustering, the LLMs are non-deterministic and can hallucinate~\cite{Sanderson2023-gj}. It should be noted that GPT-4 performs the highest in tasks related to summary and categorization~\cite{katz-gpt4}, which is how it was used in our study. To further address this concern in our work, we spot-checked the categories and topics but did not find it necessary to modify them from their original results. The topic clustering was performed by BERTopic, the LLMs were only used to attribute human-friendly meaning to the cluster results that BERTopic found. 

\subsubsection{\textbf{External Threats}} Our analysis only includes data from Stack Overflow. Although Stack Overflow is the most visited question-and-answer site for developers, it is not fully inclusive of developer thoughts. Other community forums exist, such as Server Fault, the Kubernetes Slack server, and other disparate communities on Discord and message forums. Another external threat is that search engine referral traffic to Stack Overflow could alter the reliability of \textit{views} as a measure of interest or popularity metric. 

This analysis is also predicated on the idea that Stack Overflow is effectively moderated, which we believe it to be (at least for the benefit of macro-analysis and trend analysis). Therefore, we operated under the assumption that posts were properly tagged and deduplicated. Given the large sample size of 35,714 posts, sporadic moderation errors would not significantly alter our study's results. 

\section{Related Work} \label{sec:threats}

While Kubernetes was originally developed in 2014, it did not receive much attention from researchers until 2021 and 2022. Therefore, the academic landscape of Kubernetes is limited and is still in its infancy.

During its early years, cloud computing faced many challenges, like automated server provisioning, traffic management, and data security, as identified in Zhang et al.~\cite{Zhang2010-cq}. These challenges were validated by Burns et al.~\cite{Burns2016-rr}, who proposed solving these challenges by building Kubernetes within Google and later open-sourcing and donating it to CNCF.

Several studies have looked at the current security status of Kubernetes by analyzing open-source manifests. Rahman et al.~\cite{Rahman2023-xm} searched open-source Kubernetes manifests to find common security misconfigurations. The authors found that 58\% of security misconfigurations are insecure http, 16\% are hard-coded secrets, and 10\% are absent security contexts. 

Bose et al.~\cite{Bose2021-pz} conducted a qualitative analysis study on open-source Kubernetes misconfigurations. The authors analyzed 5,193 commits and found that security misconfigurations in Kubernetes go dramatically under-reported and pose significant industry risk. Nine of the 38 repositories in their study had at least one security defect currently exposed publicly.

The security challenges of integrating Kubernetes into your CI/CD process were collected by Shevchuk et al.~\cite{Shevchuk2023-dn}. Helm, the primary package manager for Kubernetes, is used by 63\% of practitioners\cite{Zerouali2023-ay} as the preferred way to deploy applications in Kubernetes through a configuration-as-code paradigm. Blaise et al.~\cite{Blaise2022-ia} analyzed security concerns targeting Kubernetes through Helm as an attack vector. In another study, researchers mined 9,482 Helm charts\footnote{``Chart'' is the term used by helm to define a packaged (\texttt{.tar} format) set of configuration files that deploy an application into a Kubernetes cluster} to analyze the metadata and security risks by Zerouali et al.~\cite{Zerouali2023-ay}. They found that more than half of the charts were outdated and that 88\% were exposed to vulnerabilities, ultimately jeopardizing 93.7\% of the charts.  

Rahman et al.~\cite{Rahman2023-xm} conducted a gray literature review of common security practices for Kubernetes on 11 topics that are frequently discussed in online blogs and training media. The study found that Authentication and Authorization was the most popular topic, followed by Kubernetes-specific security policies, vulnerability scanning, secure logging, namespace separation, and encrypted \texttt{etcd} store. This study only addressed the training material available for Kubernetes, but does not consider the discussions happening by developers and administrators themselves. Our study seeks to expand this analysis by considering the types of questions and concerns currently happening in the community to understand areas for improvement or lack of knowledge and documentation.

Stack Overflow mining and analysis is a useful tool for understanding community opinion across various technologies. To the best of our knowledge, this research is the first to conduct such a study specific to Kubernetes, but other papers have performed similar work in adjacent categories. Barua et al.~\cite{Barua2014-di} performed an analysis of Stack Overflow to get a general overview of the topics developers are talking about across all technology sectors. They used cluster topic modeling to find overarching topics and provided detailed explanations for data cleansing Stack Overflow content. Tanzil et al.~\cite{Tanzil2023-if} analyzed 174k posts related to DevOps to understand the landscape of \mbox{DevOps} technologies, combined with a developer survey. The authors found that 13.1\% of all \mbox{DevOps} posts are related to Kubernetes, but the paper focused on general \mbox{DevOps} topics and stopped at categorizing posts as ``Kubernetes''. That paper never explores the topics \textit{within} the Kubernetes category, which our paper does.

\section{Conclusion} \label{sec:conclusion}

The rapid adoption and growth of Kubernetes across the software development landscape highlights its significance in orchestrating containerized applications at scale. However, this platform dependence has also introduced new complex security challenges. These challenges are evidenced by the discussions and concerns raised within the developer community on platforms like Stack Overflow. Our analysis of these discussions has uncovered key insights into the prevalent security issues facing Kubernetes practitioners today.

Prominently, topics such as TLS/SSL Certificate Management and Secrets Management have emerged as critical areas of focus in the evolution of the platform. These discussions underscore the need for robust mechanisms to secure communications within Kubernetes clusters and protect sensitive information. Equally, the attention towards Identity and Access Management (IAM) emphasizes the need for effective access control measures to prevent unauthorized access and potential security breaches.

Interestingly, the evolution of security-related discussions over time presents a dual narrative: while the number of new security posts has remained relatively stable, the surge in views indicates a growing interest and concern about Kubernetes security. This contrast may reflect a maturing platform where foundational security questions have been addressed, yet the demand for information and guidance continues to rise alongside Kubernetes' adoption.

These discussions highlight the community's efforts in seeking solutions while also pointing out areas requiring further clarity and resources. As Kubernetes continues to play a pivotal role in modern software infrastructure, our research encourages a healthy evolution of the platform by addressing these security concerns through continued research, improved documentation, platform features, and tooling.

\section{Data Availability} \label{sec:data}
The data and artifacts of this study are available here - https://figshare.com/s/8f9339022f81e1835dbc

\bibliographystyle{abbrv}
\bibliography{xreferences}

\end{document}